\documentclass[aps,prl,superscriptaddress,twocolumn]{revtex4-2}

\usepackage{amsmath,amssymb,mathrsfs}
\usepackage{bm}
\usepackage{braket}

\usepackage{graphicx}

\usepackage{verbatim}
\usepackage{import}
\usepackage{pdfpages}

\makeatletter
\AtBeginDocument{\let\LS@rot\@undefined}
\makeatother

\usepackage{pgffor}

\usepackage{xcolor}

\usepackage[hidelinks]{hyperref}
\hypersetup{
  colorlinks=true,
  linkcolor=blue,
  urlcolor=blue,
  citecolor=blue
}

\begin{document}

\title{Kerr-enhanced amplification of three-wave mixing  and emergent masing regimes}

\author{Ragheed Alhyder}
\email{ragheed.alhyder@ist.ac.at}
\affiliation{Institute of Science and Technology Austria (ISTA), am Campus 1, 3400 Klosterneuburg, Austria}

\author{Rishabh Sahu}
\affiliation{Institute of Science and Technology Austria (ISTA), am Campus 1, 3400 Klosterneuburg, Austria}

\author{Johannes M. Fink}
\affiliation{Institute of Science and Technology Austria (ISTA), am Campus 1, 3400 Klosterneuburg, Austria}

\author{Mikhail Lemeshko}
\affiliation{Institute of Science and Technology Austria (ISTA), am Campus 1, 3400 Klosterneuburg, Austria}

\author{Georgios M. Koutentakis}
\email{georgios.koutentakis@ist.ac.at}
\affiliation{Institute of Science and Technology Austria (ISTA), am Campus 1, 3400 Klosterneuburg, Austria}

\begin{abstract}
Integrated optical microresonators exploiting either second--order ($\chi^{(2)}$) or third--order ($\chi^{(3)}$) nonlinearities have become key platforms for frequency conversion, low--noise microwave photonics, and quantum entanglement generation. 
Here, we present an analytic theory of Kerr-enhanced three-wave mixing amplification in an electro-optic microresonator with both $\chi^{(2)}$ and $\chi^{(3)}$ nonlinearities.
We demonstrate that Kerr dressing hybridizes the optical sidebands, renormalizing the $\chi^{(2)}$ couplings and detunings.
As a result the system exhibits gain in regions where analogous bare $\chi^{(2)}$ or $\chi^{(3)}$ amplifiers are subthreshold.
Time-domain Langevin simulations confirm this threshold reduction, mapping a practical design window for experiments.
\end{abstract}
\maketitle

Nonlinear optical microresonators combine ultra-high quality factors with small mode volumes, enabling efficient frequency conversion and parametric processes at milliwatt pump powers across both quadratic ($\chi^{(2)}$) and cubic ($\chi^{(3)}$) nonlinear regimes~\cite{Vahala2003,Savchenkov2004,Furst2010,Strekalov2016,Gaeta2019,Fabre2020,Frigenti2023}.
Kerr microcombs generated by four-wave mixing have emerged as chip-scale frequency synthesizers and broadband comb sources~\cite{Kippenberg2004,DelHaye2007,Kippenberg2011,Gaeta2019}. Applications include precision metrology~\cite{Suh2016,Dutt2018}, coherent communications~\cite{Pfeifle2014,MarinPalomo2017}, and microwave photonics through parametric oscillation and dissipative Kerr soliton formation~\cite{Kippenberg2018,Geng2022}. The underlying Kerr dynamics represented in self- and cross-phase modulation, modulational instability, and soliton formation, produce intensity-dependent shifts and splittings of cavity modes and drive frequency-comb generation~\cite{Kippenberg2018,Strekalov2016}. The same nonlinearity can destabilize continuous-wave operation and induce pronounced thermo-optical bistability and thermal instabilities that constrain operating windows and access to broadband comb states~\cite{Pasquazi2018,Leshem2021,Ngek2023}.

Within this landscape, $\chi^{(2)}$ three-wave mixing is a key resource for coherent microwave-to-optical transduction and parametric amplification~\cite{Andrews2014}. In cavity electro-optic devices, a driven optical mode is coupled by the Pockels effect to a microwave resonance, realizing a direct analogue of three-wave mixing amplifiers~\cite{tsang2010,tsang2011,Rueda2016,Fan2018, rueda2019}. In the resolved-sideband, phase-matched regime, conversion efficiency is determined by the electro-optic cooperativity, $C$, which compares the three-wave coupling strength to the optical and microwave decay rates. Considerable experimental progress has focused on increasing cooperativity using higher $Q$ factors, improved optical–RF overlap, and noise suppression, leading to large classical conversion efficiencies and first near-quantum-limited transducers~\cite{Holzgrafe2020,McKenna2020,Borwka2023,SahuHease2022,Sahu2023}. However, further increasing cooperativity is technically challenging and often limited by parasitic nonlinear and thermal effects.

In electro-optic devices designed around $\chi^{(2)}$ three-wave mixing, Kerr nonlinearity ($\chi^{(3)}$) is an important such effect. The Kerr nonlinearity detunes the optical sidebands from the microwave resonance, degrades phase matching, and raises the gain or conversion threshold~\cite{Zhang2021,Rueda2016,Soltani2017}. At the same time, recent experiments on hybrid Kerr-electro-optic microcombs and on platforms that naturally host both nonlinearities show that the two processes are generically present and can strongly interact~\cite{Guo2018,Bruch2020,Nie2022,song2025}. These works suggest that the Kerr nonlinearity need not be merely parasitic in electro-optic systems, and motivate a more systematic understanding of how Kerr-induced spectral reshaping can be harnessed, rather than avoided, in $\chi^{(2)}$-based devices. This pushes for a careful theoretical investigation of hybrid $\chi^{(2)}$--$\chi^{(3)}$ microresonators to exploit Kerr-induced reshaping of the optical spectrum to engineer three-wave mixing interactions that are stronger or more resonant than in purely $\chi^{(2)}$ devices. 

\begin{figure}
    \centering
    \includegraphics[width=1.\columnwidth]{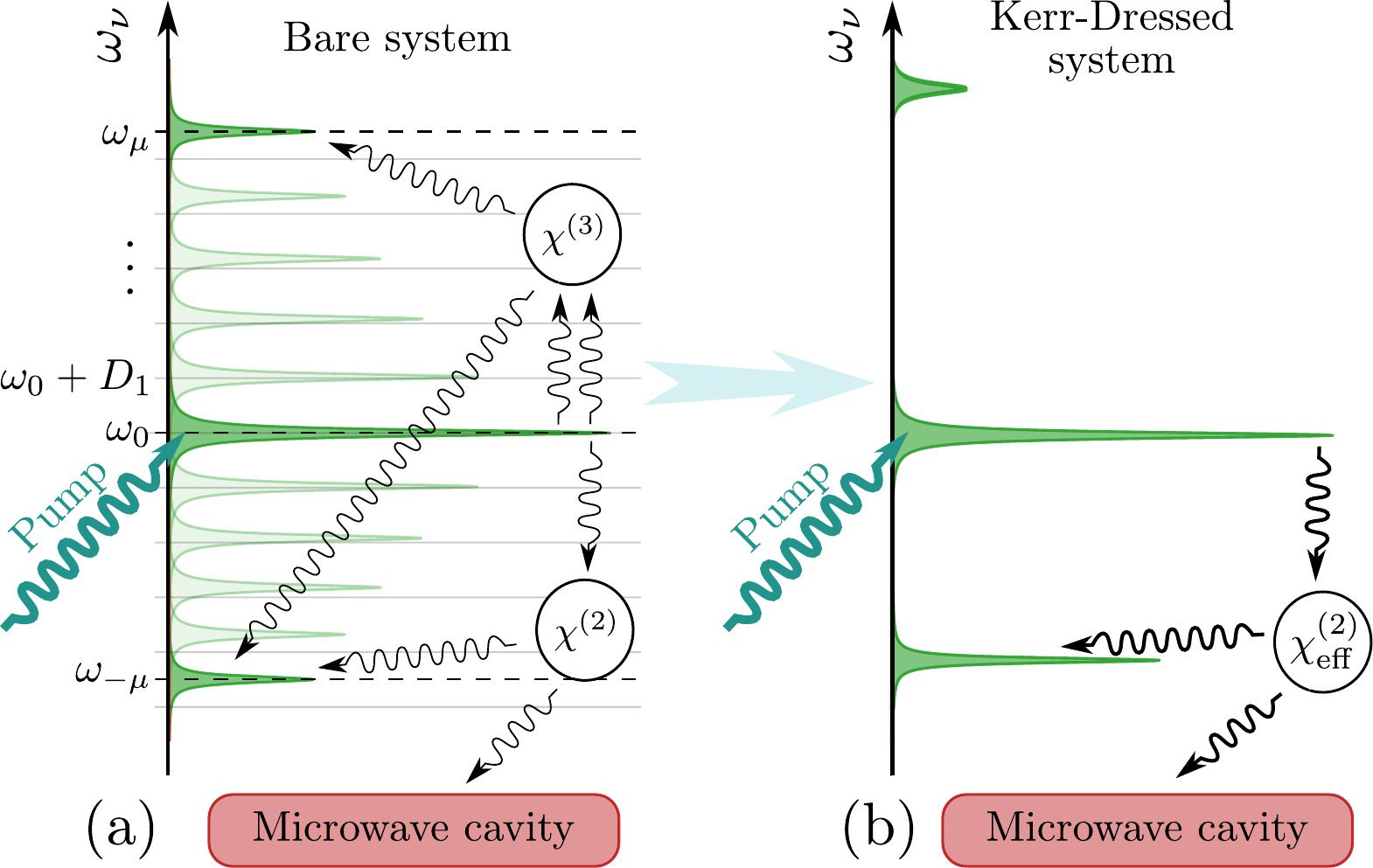}
    \caption{Description of a hybrid $\chi^{(2)}$--$\chi^{(3)}$ electro--optic resonator. (a) Optical resonances with frequencies $\omega_\nu$, $\nu = 0, \pm 1, \dots$, form an approximately equidistant ladder with spacing $D_1$ around the pumped mode $\omega_0$. The $\chi^{(2)}$ effect couples $\omega_{\nu}$ to a microwave cavity enabling up and down conversion between the $0$ and $\nu = \pm \mu$ modes by absorbing or emmiting a microwave photon. The Kerr ($\chi^{(3)}$) effect weakly couples the $0$ mode to the $\pm \nu$ sidebands by four-wave mixing. (b) We show that in the dressed-mode picture the Kerr coupling renormalizes the $\pm \mu$ modes leading to an effective 4-mode $\chi^{(2)}_{\rm eff}$ microresonator that enables parametric gain by down-converting $0$ mode photons to microwave in regimes where the individual $\chi^{(2)}$ and $\chi^{(3)}$ interactions are too weak to cause any amplification.}
    \label{fig1:Model}
\end{figure}

Here we address this question by developing a minimal yet fully analytic description of Kerr-enhanced $\chi^{(2)}$ amplification in an electro-optic microresonator. 
We show that Kerr-induced sideband hybridization yields a closed-form, intensity-tunable renormalization of both the effective detuning and the three-wave matrix elements. This produces a non-monotonic gain-threshold landscape analytically tractable by the critical electro-optic cooperativity, $C_{\mathrm{crit}}$, at which the system becomes amplifying.
We identify an extended regime with $C_{\mathrm{crit}} < 1$ where amplification is possible only within the framework of Kerr-enhanced three-wave mixing and not when either $\chi^{(2)}$ or $\chi^{(3)}$ interactions are considered in isolation. As the $\chi^{(3)}$ interaction is increased, such that the Kerr amplification regime in the absence of $\chi^{(2)}$ effect is approached, we show that $C_{\mathrm{crit}} \to 0$. This demonstrates that microwave coupling of sidebands near but below the Kerr amplification threshold can lead to enhanced three-wave mixing even for modest, $C < 1$, cooperativity values.

Starting from the standard $\chi^{(2)}$--$\chi^{(3)}$
microresonator Hamiltonian (see Supplementary materials \footnote{Supplementary materials containing the reference \cite{Rudin1976}}\nocite{Rudin1976}), we linearize around the pumped mode $a_{0}$ and obtain the dynamical matrix 
for the microwave mode $b$ and the near-resonant to it optical sidebands $a_{\pm}$~\cite{Pasquazi2018,Herr2012}. The fluctuations obey
$
  \dot{\mathbf{v}} = \mathcal{J}^{(2+3)} \mathbf{v},$ where
  $\mathbf{v} = (\delta a_{+},\,\delta a_{-}^\dagger,\,\delta b)^{\mathsf{T}},
$ is the mode's first-order fluctuations vector,
and $\mathcal{J}^{(2+3)}$ is the dynamical matrix which reads
\begin{equation}
\mathcal{J}^{(2+3)} =
\begin{pmatrix}
-\left(\dfrac{\kappa_{\rm \mu}}{2} + i \tilde\zeta_{\rm \mu} \right) & i g_{3} a_{0}^{2} & - i g_{2} a_{0} \\
- i g_{3} a_{0}^{*2} & -\left(\dfrac{\kappa_{\rm \mu}}{2} - i \tilde\zeta_{\rm \mu}\right) & i g_{2} a_{0}^{*} \\
- i g_{2} a_{0}^{*} & - i g_{2} a_{0} & -\left( \dfrac{\kappa_{\rm e}}{2} + i \zeta_{\rm e} \right)
\end{pmatrix},
\label{eq:J23_matrix}
\end{equation}
where $\tilde\zeta_{\rm \mu} = \zeta_{\rm \mu} - 2 g_{3} |a_{0}|^{2}$, is the Kerr-shifted optical sideband detuning, with $\zeta_{\rm \mu} = \omega_{\mu} - \omega_{\rm p} - D_1 \mu$ the bare detuning. Here $\omega_0$, $\omega_{\rm p}$ are the frequency of the pumped mode and pumping laser respectively, and $D_1$ is the first order dispersion coefficient, which is almost equal to the free spectral range (FSR) $D_1 \approx \omega_{\rm FSR} \equiv \omega_1 - \omega_0$ \cite{Herr2012} (see also \cite{Note1}). 
The index $\mu$ denotes the integer mode-number separation between the pumped mode and the optical sideband that is nearly resonant with the microwave cavity, see Fig.~\ref{fig1:Model}(a).
The microwave detuning is $\zeta_{\rm e}$, $\kappa_{\rm \mu}$ and $\kappa_{\rm e}$ are the total optical and microwave linewidths, respectively, and $g_{2}$ and $g_{3}$ are the three-wave and four-wave mixing coupling strengths. Here, $a_{0}$ is the steady-state amplitude of the pumped optical mode, which depends on the pump power and detuning (see \cite{Note1} for details). 

\begin{figure}
    \centering
    \includegraphics[width=1.\columnwidth]{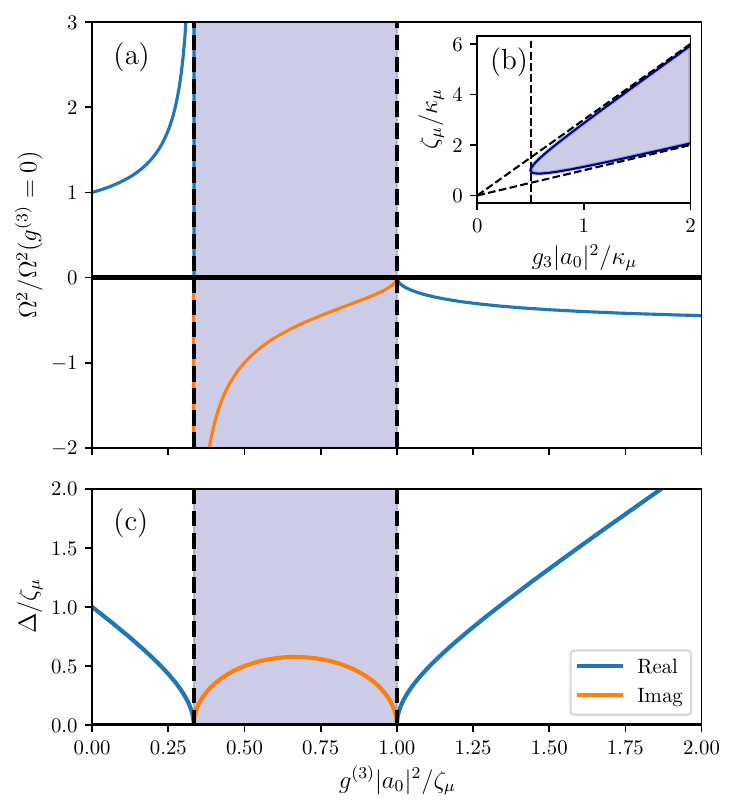}
    \caption{
        Kerr-induced renormlization of the three-wave mixing parameters.
        (a) Real and imaginary parts of the normalized coupling matrix elements
        $\Omega_{\rm eff} / \Omega_{\rm eff}\!\big(g_{3}=0\big)$ as a function of the
        Kerr-induced frequency shift
        $g_{3}|a_0|^2/\zeta_{\rm \mu}$.
        (b) Kerr-only amplification lobe in the
        $(g_3|a_0|^2/\kappa_{\rm \mu},\;\zeta_{\rm \mu}/\kappa_{\rm \mu})$ plane obtained by the
        condition ${\rm Re}(\lambda_{g_3})>0$.
        The filled area is bound by the dashed straight lines $\zeta_{\rm \mu} > g_3|a_0|^2$, $\zeta_{\rm \mu} < 3g_3|a_0|^2$  and $g_3|a_0|^2/\kappa_{\rm \mu} > 1/2$.
        (c) Real and imaginary parts of the effective detuning
        $\Delta/\zeta_{\rm \mu}$.
        The shaded region $1/3 < g_{3}|a_0|^2/\zeta_{\rm \mu} < 1$ in (a) and (b) marks the region where Kerr-induced amplification is possible (see also(b)).}
  \label{fig2:KerrDressing}
\end{figure}

Note here that the idler and signal modes $a_+ \equiv a_{\mu}$ and $a_- \equiv a_{- \mu}$ respectively, can in refer to any pair of optical modes. 
The determining factor for $\mu$ is the ratio of the frequency of the microwave mode over $D_{1}$ determining the near-resonant $\chi^{(2)}$ process. Note here that in order to achieve optical-microwave coupling, phase matching conditions should be ensured \cite{boyd2008,Rueda2016}. Strictly speaking Eq.~\eqref{eq:J23_matrix} holds only for up to quadratic dispersion of the mode frequencies,
although our methodology can be extended to higher-order dispersion terms (not shown here for brevity).

For vanishing three-wave mixing terms ($g_{2}=0$), the dynamical matrix reduces to an optical Kerr block $\mathcal{J}^{(3)}$ (first two columns and rows of Eq.~\eqref{eq:J23_matrix}) and a decoupled microwave mode. 
We can connect this block to the full three-mode problem in Eq.~\eqref{eq:J23_matrix} by taking the similarity transformation that diagonalizes the optical block $\mathcal{J}^{(3)}$ , and apply it to the full $\mathcal{J}^{(2 + 3)}$ matrix. In the new basis, the two optical modes are replaced by the Kerr eigenmodes, while the microwave amplitude is left unchanged.
This transformation leads to the effective three-wave mixing dynamical matrix
\begin{equation}
\mathcal{J}^{(2+3)}_{\rm tr} = \left(
    \begin{array}{c c c}
    - \left(\frac{\kappa_{\rm \mu}}{2} + i \Delta\right) & 0 & -i g_2 a_0 \Omega_-\\
    0 & - \left( \frac{\kappa_{\rm \mu}}{2} - i \Delta \right) & i g_2 a_0^* \Omega_+\\
    - i g_2 a_0^* \Omega_+ & %
        -i g_2 a_0 \Omega_- & %
            - \left( \frac{\kappa_e}{2} + i \zeta_{\rm e} \right)
    \end{array} \right),
\end{equation}
where $\Delta \equiv \sqrt{\Delta_{+}\Delta_{-}}$, with $\Delta_{\pm} \equiv \tilde\zeta_{\rm \mu}\pm g_3|a_0 |^2$, denotes the effective detuning and $\Omega_{\pm} \equiv \pm \sqrt{\frac{\Delta}{2 g_3 |a_0|^2}}\left(1 \pm \frac{\Delta_+}{\Delta}\right)$ are the effective coupling parameters.
This transformation greatly simplifies the treatment of the coupled $\chi^{(2)}$--$\chi^{(3)}$ system as $\mathcal{J}^{(2+3)}_{\rm tr}$ has an identical characteristic polynomial to a $\chi^{(2)}$ system following the dynamical matrix $\mathcal{J}^{(2+3)}$ for $g_3 = 0$, with the replacements $\tilde{\zeta}_{\rm \mu} \to \Delta$ and $g_2^2 |a_0|^2 \to \Omega_{\rm eff} \equiv g_2^2 |a_0|^2\Omega_{+} \Omega_- = g_2^2 |a_0|^2(\Delta_+ / \Delta)$.
Notice that for $g_3 \to 0$ we have $\Delta, \Delta_{\pm} \to \zeta_{\rm \mu}$ and recover the bare three-wave mixing parameters.

This construction makes it natural to see the off-diagonal elements as Kerr-dressed three-wave coupling parameters $\chi^{(2)}_{\rm eff}$, in addition to a single effective optical detuning $\Delta$, which plays the role of the sideband detuning in the Kerr-free problem, see Fig.~\ref{fig1:Model}(b). 
Figure~\ref{fig2:KerrDressing} illustrates how these quantities modify as a function of Kerr strength $g_{3}|a_{0}|^{2}$ to the bare detuning $\zeta_{\rm \mu}$, where (a) shows the real and imaginary parts of $\Omega_{\rm eff}$ normalized to the bare $\chi^{(2)}$ value, and (c) shows the corresponding behaviour of the dressed detuning $\Delta/\zeta_{\rm \mu}$. For weak Kerr interaction one finds $\Delta\approx \zeta_{\rm \mu}$ and $\Omega_{\rm eff}\approx g_2 |a_0 |$, so the system behaves as an almost ideal three-wave mixer with perturbatively shifted parameters. 

As the Kerr coupling $g_3 |a_0|^2$ grows, the dressed detuning $\Delta$ decreases and the effective coupling is enhanced, reaching a resonance where $\Omega_{\rm eff}$ diverges at $g_{3}|a_{0}|^{2}/\zeta_{\rm \mu}=1/3$ and $\Delta$ vanishes. Beyond this point $\Delta$ becomes purely imaginary in the interval $\zeta_{\rm \mu}/3<g_{3}|a_{0}|^{2}<\zeta_{\rm \mu}$, signalling that the Kerr nonlinearity leads to the modification of the sidebands' loss rates. This interval also corresponds to the regime where Kerr-only amplification is possible, provided $g_3 |a_0|^2/\kappa > 1/2$, as shown in Fig.~\ref{fig2:KerrDressing}(b). In this range the three-wave coupling $\Omega_{\rm eff}$ becomes imaginary, this analytic continuation to the complex plane of the response of the model implies behavior beyond what is possible by a bare $\chi^{(2)}$ electro-optic resonator. x

For $g_3 |a_0|^2 = \kappa_{\rm \mu}$ the effective coupling vanishes, $\Omega_{\rm eff} = 0$, as in this case the signal and idler bands are equally separated from the pumped mode. As a consequence, the microwave generating $\chi^{(2)}$ transport from $a_0$ to $a_-$ exactly cancels the microwave consuming $a_0$ to $a_+$ transfer process, as both mechanisms are equally probable. For even larger Kerr shifts both $|\Delta|$ and $|\Omega_{\rm eff}|$ increase, but the real part of $\Omega_{\rm eff}$ is suppressed compared to its bare value, indicating that the effective $\chi^{(2)}$ process is off resonant and becomes progressively less efficient. The negative value of $\Omega_{\rm eff}$ corresponds to a change or role of signal and idler within the effective $\mathcal{J}^{(2+3)}_{\rm tr}$ description.

\begin{figure}
    \centering
    \includegraphics[width=1.\columnwidth]{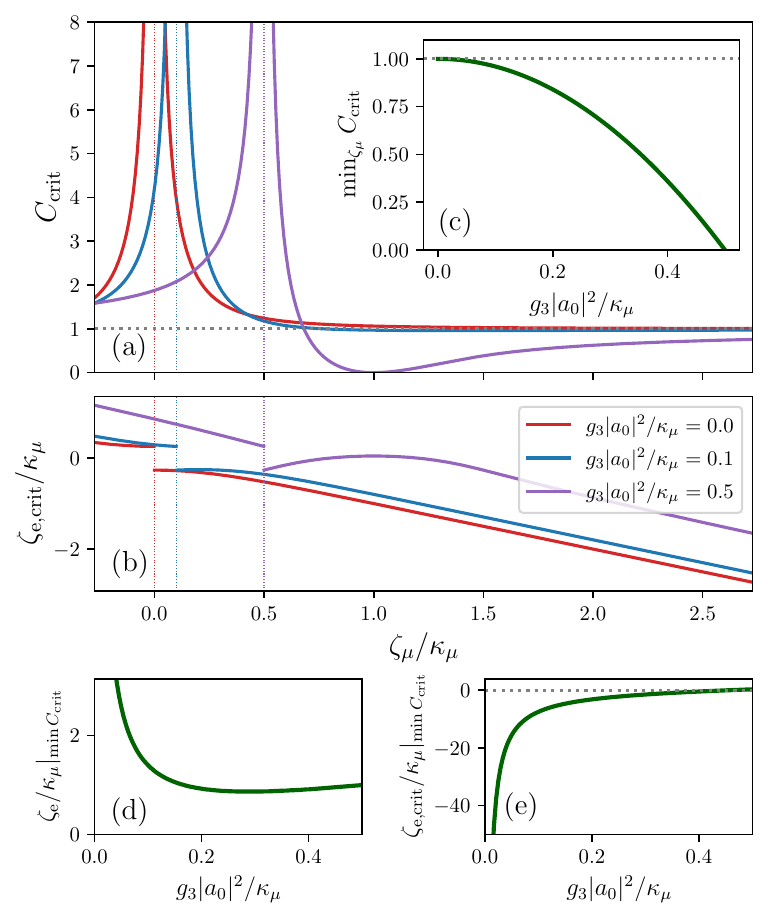}
    \caption{
    Critical cooperativity for $\chi^{(2)}$ amplification in the presence of Kerr nonlinearity. (a) The critical cooperativity $C_{\mathrm{crit}}$ and (b) the corresponding critical microwave detuning $\zeta_{{\rm e},\mathrm{crit}}/\kappa_{\rm \mu}$, obtained from Eq.~\eqref{eq:Ccrit_full_SM}, as functions of the optical detuning $\zeta_{\rm \mu}/\kappa_{\rm \mu}$ for several Kerr shifts, $g_{3}|a_0|^2/\kappa_{\rm \mu}$ (colors). In (a) the horizontal dotted line marks $C_{\mathrm{crit}}=1$, and vertical dotted lines in (a) and (b) indicate $\zeta_{\mu}= g_3 |a_0|^2$ for each Kerr strength. (c) Minimum value of critical cooperativity $\min_{\zeta_{\rm \mu}} C_{\mathrm{crit}}$, and associated (d) optical, $\zeta_{\rm \mu}/\kappa_{\rm \mu}$, and (e) microwave, $\zeta_{{\rm e},\mathrm{crit}}/\kappa_{\rm \mu}$, detuning where this minimum is attained versus $g_{3}|a_0|^2/\kappa_{\rm \mu}$.
    }
    \label{fig3:CritCoop_Detuning}
\end{figure}

From a physical point of view, this shows that the combined $\chi^{(2)}$--$\chi^{(3)}$ dynamics is not simply a competition between two independent nonlinearities. By first diagonalizing the optical Kerr block, we create hybrid modes whose frequencies and linewidths depend on $a_{0}$ and $g_{3}$. The three-wave process therefore probes these hybrid modes rather than the bare sidebands. Depending on the operating point, the Kerr shift can move one of the hybrid modes closer to the microwave resonance while keeping the other one sufficiently detuned, which enhances the relevant off-diagonal matrix elements by $\Omega_{\pm}$ and allows the amplifier to reach gain for a smaller bare $\chi^{(2)}$ coupling (corresponding to a lower pump power). If the Kerr shift becomes too strong, both hybrid modes are pushed away from the microwave resonance and the same mechanism suppresses $\Omega_{\pm}$, thereby increasing the required pump power and ultimately destroying amplification. The nonmonotonic behaviour of $\Omega_{\rm eff}$ in Fig.~\ref{fig2:KerrDressing} is thus a direct manifestation of how the Kerr interaction reshapes the underlying linear spectrum to either assist or inhibit three-wave amplification.

Since the $\mathcal{J}^{(2+3)}_{\rm tr}$ system is an effective three-wave mixing amplifier, we use the electro-optic cooperativity
$C \equiv 4 g_2^2 |a_0|^2 / (\kappa_{\rm \mu} \kappa_{\rm e})$, which quantifies
the ratio of the three-wave coupling to the optical and microwave loss rates to characterize its parametric gain and loss regimes.
This quantity in the case that a single optical mode is coupled to the microwave resonator yields the onset of amplification when $C \gtrsim 1$  \cite{tsang2010, tsang2011}. As we show, the Kerr effect redefines the critical cooperativity at which the system becomes amplifying.

Practically, we evaluate the characteristic polynomial of
$\mathcal{J}^{(2+3)}_{\rm tr}$ and impose the condition that the real part of the maximum eigenvalue is ${\rm Re}(\lambda_{\rm max}) = 0$ at its global maximum with respect to $\zeta_e$, $\partial_{\zeta_{\rm e}}\mathrm{Re}(\lambda_{\max})=0$ and solve for $C_{\rm crit}$, $\zeta_{e, {\rm crit}}$ and the imaginary part of the eigenvalue, $\lambda_{\rm im}$ and its derivative \cite{Note1}.
This yields a closed-form expression for the critical cooperativity
\begin{equation}
C_{\mathrm{crit}}(\zeta_{\mu})
=\left|
\frac{2 \lambda_{\rm im}}{3} \frac{
\bigl(\kappa_{\rm \mu}^{2}-4\Delta^{2}
+\mathcal{R}\bigr)}
{\kappa_{\rm \mu}^{2}\Delta_+}\right|,
\label{eq:Ccrit_full_SM}
\end{equation}
where
$\mathcal{R}\equiv
\sqrt{\kappa_{\rm \mu}^{4}+4\kappa_{\rm \mu}^{2}\Delta^{2}+16\Delta^{4}}$ and 
$\lambda_{\rm im} \equiv
{\rm sign} (\Delta_{+}) \sqrt{4\Delta^{2}-\kappa_{\rm \mu}^{2}+2\,\mathcal{R}}/(2 \sqrt{3})$.
A similar equation holds for the optimal detuning $\zeta_{e,\mathrm{crit}}(\Delta)$ as a function of the Kerr-dressed detuning $\Delta$ \cite{Note1}.

Figure~\ref{fig3:CritCoop_Detuning} summarizes these results. In
Fig.~\ref{fig3:CritCoop_Detuning}(a) the critical cooperativity is plotted as a
function of the bare sideband detuning $\zeta_{\rm \mu}$ for several values of the
normalized Kerr strength $g_{3}|a_0|^{2}/\kappa_{\rm \mu}$. For each curve,
$C_{\mathrm{crit}}$ diverges at
$\zeta_{\rm \mu} = g_{3}|a_0|^{2}$, where since $\Omega_{\rm eff} = 0$, the three-wave amplification is
impossible, see Fig.~\ref{fig3:CritCoop_Detuning}(a). On either side of this point the threshold decreases towards a
finite asymptotic value. As the Kerr strength $g_3 |a_0|^2$ is increased, the minimum of
$C_{\mathrm{crit}}$ decreases, see Fig.~\ref{fig3:CritCoop_Detuning}(c). 
The global minimum of the critical cooperativity with respect to both $\zeta_{e}$ and $\zeta_{\mu}$ remains below unity for all Kerr strengths, i.e. $\min_{\zeta_{\rm e},\zeta_{\mu}} C_{\mathrm{crit}}<1$. In other words, this optimum is reduced below its $g_3=0$ value, implying that Kerr dressing enables $\chi^{(2)}$ amplification in parameter regimes where a bare $\chi^{(2)}$ device cannot reach the amplification threshold. Finally, notice that for $g_3 |a_0|^2 = \kappa_{\rm \mu}/2$, the value $\min_{\zeta_{\rm e}, \zeta_{\rm \mu}} C_{\mathrm{crit}} = 0$, which is caused by the fact that the sidebands amplify by $\chi^{(3)}$ alone in this regime, see also Fig.~\ref{fig2:KerrDressing}(c). Thus the amplification observed here is a genuine cooperative effect of $\chi^{(2)}$--$\chi^{(3)}$ as neither non linearity can produce gain $C<1$ gain in the $g_3 |a_0| < \kappa_{\mu}$ regime.

In Fig.~\ref{fig3:CritCoop_Detuning}(b), we plot the corresponding optimal microwave detuning $\zeta_{e,\mathrm{crit}}/\kappa_{\rm \mu}$, which shows that the Kerr nonlinearity red-detunes the optimal operating point as the Kerr strength is increased.  In Fig.~\ref{fig3:CritCoop_Detuning}(d, e), we examine the critical values of the optical and microwave detuning where the amplification with $C_{\mathrm{crit}} < 1$ occurs. For large values of $g_3 |a_0|^2/\kappa_{\rm \mu}$ the detunings are small which might be advantageous for experimental designs implementing Kerr-enhanced $\chi^{(2)}$ amplification given also the large reduction of $C_{\rm crit}$ in this regime.

\begin{figure}
    \centering
    \includegraphics[width=1.0\columnwidth]{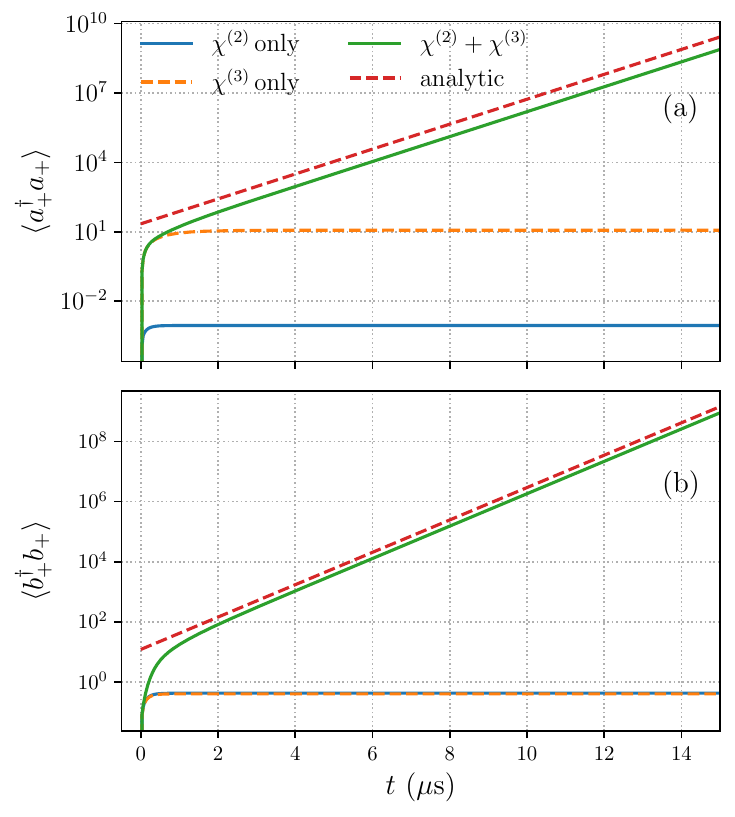}
    \caption{
    Time-domain verification of Kerr-enhanced $\chi^{(2)}$ amplification by the linearized quantum Langevin equations. (a) Signal-sideband population $\langle a_+^\dagger a_+ \rangle$ and (b) microwave population $\langle b_+^\dagger b_+ \rangle$ versus time. Three cases are considered: ($\chi^{(2)}+\chi^{(3)}$) both $C = 0.1$ and $g_3 |a_{0}|^{2}/\kappa_{\mu} = 0.49$, ($\chi^{(3)}$) $C = 0$ and $g_3 |a_{0}|^{2}/\kappa_{\mu} \neq 0.49$ and ($\chi^{(2)}$) $C = 0.1$ $g_3 = 0$. The exponential growth rate of ($\chi^{(2)}+\chi^{(3)}$) matches the analytically expected $2 {\rm Re}(\lambda_{\rm max})$ indicated by a guide-to-the-eye line. The remaining parameters are $\zeta_{\mu}/\kappa_{\mu} = 1$, $\zeta_{e} = 0$, $\kappa_{\rm \mu}/2\pi = 11~\mathrm{MHz}$, $\kappa_{\rm e}/2\pi = 1.1~\mathrm{MHz}$ and $\eta_{\rm \mu} = \eta_{\rm e} = 0.2$. Optical baths are initialized in vacuum and the internal microwave bath has one thermal photon.}
    \label{fig4:TimeDomainSim}
\end{figure}

Kerr dressing reshapes the effective $\chi^{(2)}$ amplifier in a non-monotonic way (Fig.~\ref{fig3:CritCoop_Detuning}). While at weak Kerr strength the impact of $\chi^{(3)}$ is minimal necessitating large $\zeta_e$, $\zeta_{\mu}$ detunings and strong $C>1$ for amplification, at intermediate Kerr strength, optical hybridization pulls one mode toward the microwave resonance, enhancing the effective three-wave coupling and yielding a significant reduction of $C_{\rm crit}$.
At larger Kerr shifts the $\chi^{(3)}$ gain stemming from ${\rm Im}(\Delta)$ increases resulting to the Kerr modulational instability. Optimal
Kerr-enhanced three-wave mixing is obtained by tuning the microwave mode near a signal–idler sideband pair just below the Kerr amplification threshold.

To verify that the Kerr-assisted enhancement is realized for realistic
parameters, we solve the linearized quantum Langevin equations in the time
domain (see \cite{Note1} for details) using experimentally motivated parameters, see Fig.~\ref{fig4:TimeDomainSim}.

We compare a $\chi^{(2)}$--$\chi^{(3)}$ system where the bare electro-optic cooperativity is subthreshold, $C=0.1$, and set $g_{3} |a_0|^2 =0.49 \kappa_{\mu}$ close to the Kerr amplification threshold. Both the sideband $\langle a^{\dagger}_{+}a_{+} \rangle$, Fig.~\ref{fig4:TimeDomainSim}(a), and the microwave $\langle b^{\dagger}_{+}b_{+} \rangle$, Fig.~\ref{fig4:TimeDomainSim}(b), populations experience gain $\propto e^{2 {\rm Re}(\lambda_{\rm max}) t}$ predicted by $\mathcal{J}_{\rm tr}^{(2+3)}$. In contrast, if we set either $C=0$ or $g_3 = 0$ both populations saturate to a bounded value, demonstrating that the gain is indeed a synergistic phenomenon of the nonlinearities.

In summary, we have shown that the interplay of $\chi^{(2)}$ and $\chi^{(3)}$ nonlinearities in a electro-optic resonator can lead to a Kerr-assisted enhancement of three-wave mixing processes. By diagonalizing the Kerr-induced optical block, we identified Kerr-dressed hybrid modes that modify the effective three-wave coupling and detuning, enabling amplification at lower cooperativities than in the pure $\chi^{(2)}$ case. Our analytical expressions for the critical cooperativity and detuning provide a design map for optimizing such hybrid nonlinear devices.
This work opens several avenues for future research. Experimentally, the predicted Kerr-assisted enhancement could be explored in integrated photonic platforms combining strong $\chi^{(2)}$ and $\chi^{(3)}$ nonlinearities, such as lithium niobate or silicon nitride resonators.

\vspace{0.1cm}
\begin{acknowledgments}
R.\ A. acknowledges funding from the Austrian Academy of Science ÖAW grant No. PR1029OEAW03.
This research was funded in whole or in part by the Austrian Science Fund (FWF) [10.55776/F1004]. For open access purposes, the author has applied a CC BY public copyright license to any author accepted manuscript version arising from this submission.
This work was in part supported by the European Research Council under grant agreement no. 101089099 (ERC CoG cQEO), and 101248662 (ERC POC CoupledEOT).
\end{acknowledgments}

\bibliographystyle{apsrev4-2}
\bibliography{bibliography}

\newpage
\clearpage
\onecolumngrid


\section*{Supplementary material}

# Supplementary Material for "Kerr-enhanced three-wave mixing amplification and emergent masing regimes"

Ragheed Alhyder,[1] Rishabh Sahu,[1] Johannes M. Fink,[1] Mikhail Lemeshko,[1] and Georgios M. Koutentakis[1]

[1]*Institute of Science and Technology Austria (ISTA), am Campus 1, 3400 Klosterneuburg, Austria*

## A. Symbol table

In Table I we explicate the conventions used in both the main text and supplementary materials. Optical modes are indexed by an integer $\mu$ relative to the pumped mode $\mu = 0$. We focus on one nearly resonant sideband pair $\pm \mu$ and one microwave mode near the $\mu$-th FSR.

| Symbol | Relation | Reference | Description |
|---|---|---|---|
| $a_0$ | input | Sec. I A | steady-state amplitude of pumped optical mode |
| $a_\nu$ | output | Sec. I A | optical sidebands $\nu \in \mathbb{N}$, $\nu \neq 0$ modes away from the pumped |
| $a_+$ | $a_+ \equiv a_{+\mu}$, output | Sec. I C | near microwave resonant signal sideband |
| $a_-$ | $a_+ \equiv a_{+\mu}$, output | Sec. I C | near the microwave resonance idler sideband |
| $b$ | output | Sec. I A | intracavity microwave mode amplitude (rotating frame) |
| $\omega_{\rm p}$ | input | Sec. I A | pump laser frequency driving $\nu = 0$ mode |
| $\omega_\nu$ | $\omega_\nu = \omega_0 + \sum_{n=1} \frac{1}{n!} D_n \nu^n$ | Sec. I A | optical resonance frequency of mode $\mu$ |
| $D_1$ | input | Sec. I A | first-order dispersion coefficient ($D_1 \simeq \omega_{\rm FSR}$) |
| $D_n$ | input | Sec. I A | higher-order dispersion coefficients |
| $\omega_e$ | input | Sec. I A | microwave cavity resonance frequency |
| $\kappa_\mu$ | $\kappa_\mu = \kappa_{\mu,\rm in} + \kappa_{\mu,\rm ex}$ | Sec. I A | total optical linewidth of sideband modes |
| $\kappa_e$ | $\kappa_e = \kappa_{e,\rm in} + \kappa_{e,\rm ex}$ | Sec. I A | total microwave linewidth |
| $\kappa_{r,\rm in}$ | input | Sec. I A | intrinsic loss rate for the resonator $r = \{e, \mu\}$ |
| $\kappa_{r,\rm ex}$ | input | Sec. I A | extrinsic loss rate for the resonator $r = \{e, \mu\}$ |
| $\eta_r$ | $\eta_r = \kappa_{r,\rm ex}/\kappa_r$ | Sec. IV B | extrinsic coupling ratio for the resonator $r \in \{e, \mu\}$ |
| $g_2$ | input | Sec. I A | effective three-wave ($\chi^{(2)}$) coupling rate |
| $g_3$ | input | Sec. I A | effective Kerr ($\chi^{(3)}$) coupling rate |
| $\zeta_\mu$ | $\zeta_\mu = \omega_\mu - \omega_{\rm p} - D_1 \mu$ | Sec. I A | optical detuning in the chosen rotating frame |
| $\zeta_0$ | $\zeta_0 = \omega_0 - \omega_{\rm p}$ | Sec. I A | pump detuning of the driven mode |
| $\zeta_e$ | $\zeta_e = \omega_e - \mu D_1$ | Sec. I A | microwave detuning in the chosen rotating frame |
| $\tilde{\zeta}_\mu$ | $\tilde{\zeta}_\mu = \zeta_\mu - 2g_3\|a_0\|^2$ | Sec. I C | Kerr-shifted sideband detuning |
| $\Delta_\pm$ | $\Delta_\pm = \tilde{\zeta}_\mu \pm g_3\|a_0\|^2$ | Sec. II A | Kerr-dressed mode overlap parameters |
| $\Delta$ | $\Delta = \sqrt{\Delta_+ \Delta_-}$ | Sec. II A | Kerr-dressed mode detuning |
| $\Omega_\pm$ | $\Omega_\pm = \pm\sqrt{\frac{\Delta}{2g_3\|a_0\|^2}}\left(1 \pm \frac{\Delta_+}{\Delta}\right)$ | Sec. II B | Kerr-modified electro-optical couplings in the effective $\chi^{(2)}$-only picture |
| $\Omega_{\rm eff}$ | $\Omega_{\rm eff} = g_2\|a_0\|^2 \Omega_+ \Omega_-$ | Sec. II B | Kerr-modified effective $\chi^{(2)}$ interaction parameter |
| $C$ | $C = \frac{4g_2^2\|a_0\|^2}{\kappa_\mu \kappa_e}$ | Sec. III | Bare (single-mode) electro-optical cooperativity |
| $C_{\rm crit}$ | output | Sec. III | Critical cooperativity at the onset of amplification for the $\chi^{(2)}(\text{-}\chi^{(3)})$ system |

TABLE I. Symbols and their descriptions for the quantities used throughout the Supplementary materials. The inputs and outputs of our approach are marked in the column Relation. The equations that connect a derived measure to inputs or outputs are also denoted in this column.

### B. Figure–equation map (main text)

For quick reference:

a. *Fig.* 2(a) – Interaction parameter of the $\mathcal{J}_{\text{tr}}^{(2+3)}$ dynamical matrix (see Eq. (S24) and Sec. II B)

$$\Omega_{\text{eff}} = g_2^2 \, |a_0|^2 \, \Omega_+^* \Omega_- = g_2^2 \, |a_0|^2 \, \frac{\Delta_+}{\Delta} = g_2^2 \, |a_0|^2 \, \frac{\sqrt{\left(\zeta_o - 2g_3 \, |a_0|^2\right)^2 - \left(g_3 \, |a_0|^2\right)^2}}{\zeta_o - 3g_3 \, |a_0|^2}. \tag{S1}$$

b. *Fig.* 2(b) – Amplification regime for four-wave mixing (see Eq. (S20) and Sec. II A)

$$2g_3|a_0|^2 - \sqrt{(g_3|a_0|^2)^2 - \frac{\kappa_\mu^2}{4}} < \zeta_\mu < 2g_3|a_0|^2 + \sqrt{(g_3|a_0|^2)^2 - \frac{\kappa_\mu^2}{4}}. \tag{S2}$$

c. *Fig.* 2(c) – Effective detuning (see Eq. (S19) and Sec. II A)

$$\Delta = \sqrt{\Delta_+ \Delta_-} = \sqrt{\left(\zeta_\mu - 2g_3 \, |a_0|^2\right)^2 - \left(g_3 \, |a_0|^2\right)^2}. \tag{S3}$$

d. *Fig.* 3(a) – Critical three-wave mixing cooperativity (see Eq. (S31a) and Sec. III)

$$C_{\text{crit}} = \text{sign}\left(\zeta_\mu - g_3|a_0|^2\right) \frac{\sqrt{4\Delta^2 - \kappa_\mu^2 + 2\sqrt{\kappa^4 + 4\Delta^2 \kappa_\mu^2 + 16\Delta^4}}}{3\sqrt{3}\kappa_\mu^2(\zeta_\mu - g_3|a_0|^2)} \left(\kappa_\mu^2 - 4\Delta^2 + \sqrt{\kappa_\mu^4 + 4\Delta^2 \kappa_\mu^2 + 16\Delta^4}\right). \tag{S4}$$

e. *Fig.* 3(b) – Critical microwave detuning (see Eq. (S31b) and Sec. III)

$$\zeta_{e;\text{crit}} = \text{sign}\left(\zeta_\mu - g_3|a_0|^2\right) \frac{\sqrt{4\Delta^2 - \kappa_\mu^2 + 2\sqrt{\kappa_\mu^4 + 4\Delta^2 \kappa_\mu^2 + 16\Delta^4}}}{2\sqrt{3}\kappa_\mu(\kappa_\mu^2 + 4\Delta^2)} \left[\kappa_e \sqrt{\kappa_\mu^4 + 4\Delta^2 \kappa_\mu^2 + 16\Delta^4} - \kappa_\mu^3 - 4\Delta^2(\kappa_\mu + \kappa_e)\right]. \tag{S5}$$

f. *Fig.* 3(c) – Minimum critical three-wave mixing cooperativity (see Sec. III)

$$\min_{\zeta_\mu} C_{\text{crit}} = 1 - \left(\frac{2g_3|a_0|^2}{\kappa_\mu}\right)^2. \tag{S6}$$

g. *Fig.* 3(d) – Location of minimum critical three-wave mixing cooperativity in terms of optical detuning (see Sec. III)

$$\zeta_\mu\big|_{\min_{\zeta_\mu} C_{\text{crit}}} = \frac{3}{2} g_3|a_0|^2 + \frac{\kappa_\mu^2}{8g_3|a_0|^2}. \tag{S7}$$

h. *Fig.* 3(e) – Location of minimum critical three-wave mixing cooperativity in terms of optical detuning (see Sec. III)

$$\zeta_{e;\text{crit}}\big|_{\min_{\zeta_\mu} C_{\text{crit}}} = -\frac{\kappa_\mu^2}{8g_3|a_0|^2} \left[1 + \left(1 + 2\frac{\kappa_e}{\kappa_\mu}\right)\left(\frac{2g_3|a_0|^2}{\kappa_\mu}\right)^2\right]. \tag{S8}$$

# I. MODEL AND EQUATIONS OF MOTION

## A. Coupled resonator equations-of-motion

By following the usual formalism, see e.g. [S1], we arrive at the following equations of motion after the inclusion of the $\chi^{(2)}$, $\chi^{(3)}$ and noise terms

$$\begin{aligned}
\dot{a}_0 &= -\left(\frac{\kappa_\mu}{2} + i\zeta_0\right)a_0 - ig_2\Big[b\,a_{-\mu} + b^*\,a_\mu\Big] + ig_3 \sum_{\nu',\nu''} a_{\nu'}a_{\nu''}a^*_{\nu'+\nu''} \\
&\quad + \sqrt{\eta\kappa}f + \sqrt{\kappa_{\text{ex}}}\,A_{0,\text{ex}} + \sqrt{\kappa_{\text{in}}}\,A_{0,\text{in}}, \\
\dot{a}_\nu &= -\left(\frac{\kappa_\mu}{2} + i\zeta_\nu\right)a_\nu - ig_2\Big[b\,a_{\nu-\mu} + b^*\,a_{\nu+\mu}\Big] + ig_3 \sum_{\nu',\nu''} a_{\nu'}a_{\nu''}a^*_{\nu'+\nu''-\nu} \\
&\quad + \sqrt{\kappa_{\text{ex}}}\,A_{\nu,\text{ex}} + \sqrt{\kappa_{\text{in}}}\,A_{\nu,\text{in}}, \\
\dot{b} &= -\left(\frac{\kappa_e}{2} + i\zeta_e\right)b - \frac{ig_2}{2}\sum_\nu \left(a_\nu a^*_{\nu-k} + a^*_\nu a_{\nu+k}\right) \\
&\quad + \sqrt{\kappa_{e,\text{ex}}}\,B_{\text{ex}} + \sqrt{\kappa_{e,\text{in}}}\,B_{\text{in}},
\end{aligned} \quad (S9)$$

where the symbols are defined in Table I and in addition we have introduced the optical, $A$, and microwave, $B$, noise impulse terms and $f$ the external drive strength in photons per unit time of the driven mode. To eliminate the time-dependence of the $\chi^{(2)}$ and $\chi^{(3)}$ terms the rotating frame defined by

$$\begin{aligned}
a_\nu(t) &= a_{\nu,\text{lab}}(t)e^{-i(\omega_\nu - \omega_p - \nu D_1)t}, \\
b(t) &= b_{\text{lab}}(t)e^{-i(\omega_e - \mu D_1)t},
\end{aligned} \quad (S10)$$

has been adopted, where $a_{\nu,\text{lab}}(t)$, $b_{\text{lab}}(t)$ encode the full time dependence of the optical and microwave modes respectively. This rotating frame assumes that the optical modes lye in an equidistant, i.e. dispersionless frequency grid. As a consequence any deviation from outdistance is encoded by a mode dependent detuning

$$\zeta_\nu = \omega_\mu - \omega_p - \nu D_1 = \zeta_0 + \sum_{n=2}^\infty D_n \nu^n. \quad (S11)$$

Finally, the spatial dependence of the modes, encoding the move overlaps, has been absorbed in the definition of the $g_2$ and $g_3$ coupling parameters.

## B. Driven pumped mode and steady state

Throughout our work we assume that the driven mode is strongly driven such that it is populated by a macroscopic number of photons making noise sources (both thermal and quantum) irrelevant. The only physical mechanism counteracting the pulse is the loss rate and thus the steady state of the system is found by solving

$$0 = -\left(\frac{\kappa_\mu}{2} + i(\zeta_0 - g_3|a_0|^2)\right)a_0 + \sqrt{\eta_\mu \kappa_\mu}\,f. \quad (S12)$$

Solutions of this equation are well known in the literature, see e.g. [S1] and will not be repeated here. For our purposes $a_0$ is assumed constant and is a given input of our model.

## C. Linearization and dynamical matrix

To proceed we linearize the equations of motion around the pumped steady state $|a_0| \gg 1/\kappa_\mu$ and vaccuum for all other modes $a_{\nu \neq 0} = 0$ and $b = 0$. Ignoring the modes excited by noise for the time being, we keep only the nearly resonant mode triplet $(a_{+\mu}, a_{-\mu}, b)$ and define its fluctuations $a_{+\mu} \approx 0 + \delta a_+$, $a_{-\mu} \approx 0 + \delta a_-$, $b \approx 0 + \delta b$. The linearized fluctuation vector

$$v \equiv (\delta a_+,\ \delta a_-^\dagger,\ \delta b)^\mathsf{T} \quad (S13)$$

obeys

$$\dot{v} = \mathcal{J}^{(2+3)} v + \text{(noise drives)}, \tag{S14}$$

with drift (dynamical) matrix identical to Eq. (1) of the main text:

$$\mathcal{J}^{(2+3)} = \begin{pmatrix} -\left(\frac{\kappa_\mu}{2} + i\tilde{\zeta}_\mu\right) & ig_3 a_0^2 & -ig_2 a_0 \\ -ig_3 a_0^{*2} & -\left(\frac{\kappa_\mu}{2} - i\tilde{\zeta}_\mu\right) & ig_2 a_0^* \\ -ig_2 a_0^* & -ig_2 a_0 & -\left(\frac{\kappa_e}{2} + i\zeta_e\right) \end{pmatrix}, \tag{S15}$$

where the Kerr-shifted sideband detuning is

$$\tilde{\zeta}_\mu = \zeta_\mu - 2g_3|a_0|^2. \tag{S16}$$

Stability is determined by eigenvalues $\lambda_j$ of $\mathcal{J}^{(2+3)}$: the system is stable if $\max_j \Re(\lambda_j) < 0$, as in this case all noise-driven fluctuations are decaying in time. While the onset of amplification is reached when $\max_j \Re(\lambda_j) \geq 0$, as the corresponding mode if it gets thermally occupied will amplify in time.

## II. KERR DRESSING OF THE OPTICAL BLOCK AND EFFECTIVE THREE-WAVE PARAMETERS

Having outlined the background knowledge and defined our setup, we now focus on the details of our main contribution. This is namely the reduction of a $\chi^{(2)}$-$\chi^{(3)}$ transducer to an effective $\chi^{(2)}$ one and the analysis of its amplification regimes.

### A. Kerr-only block and its eigenmodes

Focussing solely on the optical modes, i.e the upper-left $2 \times 2$ block of Eq. (S15), we obtain the Kerr dynamical matrix

$$\mathcal{J}^{(3)} = \begin{pmatrix} -\frac{\kappa_\mu}{2} - i\tilde{\zeta}_\mu & ig_3 a_0^2 \\ -ig_3 a_0^{*2} & -\frac{\kappa_\mu}{2} + i\tilde{\zeta}_\mu \end{pmatrix}. \tag{S17}$$

By invoking the spectral theorem this matrix can be diagonalized as

$$\mathcal{J}^{(3)} = \underbrace{\frac{1}{2\sqrt{(\Delta_+ - \Delta_-)\Delta}} \begin{pmatrix} \Delta_+ + \Delta_- + 2\Delta & e^{i2\phi}(\Delta_+ + \Delta_- - 2\Delta) \\ e^{-i2\phi}(\Delta_+ - \Delta_-) & \Delta_+ - \Delta_- \end{pmatrix}}_{\equiv V} \begin{pmatrix} -\left(\frac{\kappa_\mu}{2} + i\Delta\right) & 0 \\ 0 & -\left(\frac{\kappa_\mu}{2} - i\Delta\right) \end{pmatrix} V^{-1}, \tag{S18}$$

with $\phi = \arg(a_0)$, $\Delta_\pm = \tilde{\zeta}_\mu \pm g_3|a_0|^2$ and $\Delta = \sqrt{\Delta_+ \Delta_-}$. Therefore, the eigenvalues read

$$\lambda_\pm^{(3)} = -\frac{\kappa_\mu}{2} \pm i\Delta, \tag{S19}$$

so the Kerr-induced amplification is triggered by modulational instability $\Im(\Delta) > \kappa_\mu/2$ leading to $\Re(\lambda_+^{(3)}) > 0$. This condition directly leads to

$$2g_3|a_0|^2 - \sqrt{(g_3|a_0|^2)^2 - \frac{\kappa_\mu^2}{4}} < \zeta_\mu < 2g_3|a_0|^2 + \sqrt{(g_3|a_0|^2)^2 - \frac{\kappa_\mu^2}{4}}, \tag{S20}$$

provided that $g_3|a_0|^2 > \kappa_\mu/2$. The behaviour of $\Delta$ as a function of $\zeta_\mu/(g_3|a_0|^2)$ is depicted in Fig. 2(b) and the instability region of Eq. (S20) is depicted in the inset of Fig. 2(a) of the main text.

In the main text we are mainly interested in the region $g_3|a_0|^2 < \kappa_\mu/2$ where Kerr-induced amplification, solely based on $\chi^{(3)}$, is impossible. As shown in the following subsection the spectral expansion of $\mathcal{J}^{(3)}$ plays a crucial role in the simplification of $\mathcal{J}^{(2+3)}$.

### B. Similarity transform and Kerr-dressed detuning

To simplify $\mathcal{J}^{(2+3)}$ we begin by defining $V_3 = \begin{pmatrix} V & 0 \\ 0 & 1 \end{pmatrix}$ and similarly for $V_3^{-1} = \begin{pmatrix} V^{-1} & 0 \\ 0 & 1 \end{pmatrix}$. Then we define the transformed $\mathcal{J}_{\text{tr}}^{(2+3)} = V_3^{-1} \mathcal{J}^{(2+3)} V_3$ which reads equivalently to Eq. (2) of the main text:

$$\mathcal{J}_{\text{tr}}^{(2+3)} = \begin{pmatrix} -\left(\frac{\kappa_\mu}{2} + i\Delta\right) & 0 & -ig_2 a_0 \Omega_- \\ 0 & -\left(\frac{\kappa_\mu}{2} - i\Delta\right) & ig_2 a_0^* \Omega_+ \\ -ig_2 a_0^* \Omega_+ & -ig_2 a_0 \Omega_- & -\left(\frac{\kappa_e}{2} + i\zeta_e\right) \end{pmatrix}, \tag{S21}$$

where the dressed couplings read

$$\Omega_\pm = \pm\sqrt{\frac{\Delta}{2g_3|a_0|^2}}\left(1 \pm \frac{\Delta_+}{\Delta}\right), \tag{S22}$$

To proceed let us compare the characteristic polynomial of $\mathcal{J}^{(2)}$ with $\mathcal{J}_{\text{tr}}^{(2+3)}$. This comparison yields

$$C\left(\lambda, \mathcal{J}^{(2)}\right) = -\lambda^3 - \left(\kappa_\mu + \frac{\kappa_e}{2} + i\zeta_e\right)\lambda^2 - \left[\frac{\kappa_\mu}{4}(\kappa_\mu + 2\kappa_e + i4\zeta_e) + \zeta_\mu^2\right]\lambda$$
$$+ i2\zeta_\mu g_2^2|a_0|^2 - \frac{1}{8}(\kappa_e + i2\zeta_e)(\kappa_\mu^2 + 4\zeta_\mu^2) \tag{S23a}$$

$$C\left(\lambda, \mathcal{J}_{\text{tr}}^{(2+3)}\right) = -\lambda^3 - \left(\kappa_\mu + \frac{\kappa_e}{2} + i\zeta_e\right)\lambda^2 - \left[\frac{\kappa_\mu}{4}(\kappa_\mu + 2\kappa_e + i4\zeta_e) + \Delta^2\right]\lambda$$
$$+ i2\Delta\Omega_{\text{eff}} - \frac{1}{8}(\kappa_e + i2\zeta_e)(\kappa_\mu^2 + 4\Delta^2). \tag{S23b}$$

The above equation explicates that the Kerr-effect leads to an effective renormalization of $\zeta_\mu \to \Delta$ (compare the third and last terms of Eq. (S23b) to (S23a)) and of the three-wave mixing coupling term $g_2|a_0|$ to

$$\Omega_{\text{eff}} \equiv g_2^2|a_0|^2\Omega_+\Omega_- = g_2^2|a_0|^2\frac{\Delta_+}{\Delta}, \tag{S24}$$

(compare the fourth terms of Eq. (S23b) to (S23a)). The behavior of $\Omega_{\text{eff}}/\Omega_{\text{eff}}(g_3 = 0)$ is depicted in Fig. 2(a) in the main text. Notice that if $\Delta$ and $\Omega_{\text{eff}}$ were always real then $\mathcal{J}_{\text{tr}}^{(2+3)}$ would be subject to the same limitations as $\mathcal{J}^{(2)}$. However, since $\Delta$ and $\Omega_{\text{eff}}$ can take complex values for $1/3 < \zeta_\mu/(g_3|a_0|^2) < 1$ the characteristic polynomial $C(\lambda, \mathcal{J}_{\text{tr}}^{(2+3)})$ is an analytic continuation of $C(\lambda, \mathcal{J}^{(2)})$ and thus exhibits new properties.

Therefore, Eq. (S21) and Eq. (S23) is the rigorous statement of Kerr-modified (and as we show enhanced) three-wave mixing: the Kerr block reshapes the optical spectrum (via $\Delta$) and renormalizes the effective $\chi^{(2)}$ matrix element (via $\Delta_+/\Delta$). This transformation paves the way to a rigorous derivation of the amplification regimes by the combined influence of $\chi^{(2)}$ and $\chi^{(3)}$ effects outlined in the following.

## III. AMPLIFICATION ONSET: CRITICAL COOPERATIVITY AND OPTIMAL MICROWAVE DETUNING

The amplification threshold can be identified by the eigenvalue with maximal real part, $\lambda$, of $\mathcal{J}^{(2)}$ or $\mathcal{J}_{\text{tr}}^{(2+3)}$ exhibiting a global maximum with a value of zero as the tunable parameters of the system are varied. This tangency criterion ensures that when above threshold the real part of the eigenvalue (corresponding to the growth/loss rate) will "cut the zero plane" allowing for $\Re(\lambda) > 0$ (i.e. growth) inside the parametric region defined by the cut.

To make this notion explicit we have to clarify the role of the system parameters. From an experimental perspective the parameters $\kappa_\mu$, $\kappa_e$, $g_2$ and $g_3$ are fixed by the fabrication or are difficult to tune for a given device. Therefore, the tunable parameters are $\zeta_\mu$, $\zeta_e$ and $a_0$. Since, $a_0$ is a measure of both the $\chi^{(2)}$ and $\chi^{(3)}$ interaction as the combinations $g_2|a_0|$ and $g_3|a_0|^2$ characterize these interactions, it cannot be interpreted as a maximization parameter. Instead, it is absorbed in the definition of the threshold of amplification. The quantity we use is the electro-optic cooperativity

$$C = \frac{4g_2^2|a_0|^2}{\kappa_\mu \kappa_e}, \tag{S25}$$

which defines the amplification threshold in terms of a critical value $C_\text{crit}$. Therefore, the optical and microwave detunings $\zeta_\mu$ and $\zeta_e$ respectively are the only candidates for the variable that we maximize the eigenvalue on. Given also the known threshold of $C_\text{crit} = 1$ for the amplification of a single mode and optimized $\zeta_e$ [S2], it makes sense to define the critical cooperativity, $C_\text{crit}$, as the minimal value of $C$ where $\max_{\zeta_e} \Re(\lambda) = 0$. Within this definition $C_\text{crit} = C_\text{crit}(\zeta_\mu)$, a functional dependence demonstrated in Fig. 3(a) of the main text.

Below we provide the general mathematical framework for evaluating $C_\text{crit}$ within this definition. Afterwards we apply the relevant procedure to calculate $C_\text{crit}$ for the case of $g_3 = 0$, $\mathcal{J}^{(2)}$, and then based on this we generalize to $\mathcal{J}_\text{tr}^{(2+3)}$.

### A. Mathematical background: An implicit differentiation technique

In general, given a matrix $M = M(\boldsymbol{q})$ that depends parametrically on $N_p$ parameters, $\boldsymbol{q} = (q_1, \ldots, q_{N_p})$, each eigenvalue viewed as a function of these parameters, $\lambda(\boldsymbol{q})$, is a root of the characteristic polynomial, $\chi_M(\lambda; \boldsymbol{q})$, for all $\boldsymbol{q}$. This means that the auxiliary function $\tilde{\chi}_M(\boldsymbol{q}) = \chi_M(\lambda(\boldsymbol{q}), \boldsymbol{q}) = 0$ for all values of $\boldsymbol{q}$ and a given eigenvalue branch $\lambda(\boldsymbol{q})$. Therefore, the partial derivatives of this identically vanishing function, $\frac{\partial \tilde{\chi}_M}{\partial q_j}$, with $j = 1, 2, \ldots, N_p$, also vanish. By applying the chain-rule, we can express $\frac{\partial \tilde{\chi}_M}{\partial q_j} = 0$ in terms of $\frac{\partial \chi_M}{\partial q_j}$, $\frac{\partial \chi_M}{\partial \lambda}$ and $\frac{\partial \lambda}{\partial q_j}$ if the partial derivatives of the eigenvalues $\frac{\partial \lambda}{\partial q_j}$ are treated self-consistently. The last step requires that the eigenvalues are distinct such that $\lambda(\boldsymbol{q})$ is continuously differentiable. These mathematical considerations when combined with complex-valued characteristic polynomial and eigenvalues give the equations

$$\begin{aligned}
&\Re\left(\chi_M(\lambda_\text{re}(\boldsymbol{q}) + i\lambda_\text{im}(\boldsymbol{q}); \boldsymbol{q})\right) = 0, \\
&\Im\left(\chi_M(\lambda_\text{re}(\boldsymbol{q}) + i\lambda_\text{im}(\boldsymbol{q}); \boldsymbol{q})\right) = 0, \\
&\Re\left(\frac{\partial \chi_M}{\partial \lambda}\frac{\partial \lambda}{\partial q_j} + \frac{\partial \chi_M}{\partial q_j}\right)\bigg|_{\lambda = \lambda_\text{re}(\boldsymbol{q}) + i\lambda_\text{im}(\boldsymbol{q}), \frac{\partial \lambda}{\partial q_j}\big|_{\boldsymbol{q}} = \frac{\partial \lambda_\text{re}}{\partial q_j} + i\frac{\partial \lambda_\text{im}}{\partial q_j}\big|_{\boldsymbol{q}}, \boldsymbol{q}} = 0, \text{ for } j = 1, 2, \ldots, N_p, \\
&\Im\left(\frac{\partial \chi_M}{\partial \lambda}\frac{\partial \lambda}{\partial q_j} + \frac{\partial \chi_M}{\partial q_j}\right)\bigg|_{\lambda = \lambda_\text{re}(\boldsymbol{q}) + i\lambda_\text{im}(\boldsymbol{q}), \frac{\partial \lambda}{\partial q_j}\big|_{\boldsymbol{q}} = \frac{\partial \lambda_\text{re}}{\partial q_j} + i\frac{\partial \lambda_\text{im}}{\partial q_j}\big|_{\boldsymbol{q}}, \boldsymbol{q}} = 0, \text{ for } j = 1, 2, \ldots, N_p,
\end{aligned} \quad (S26)$$

where $\lambda_\text{re}(\boldsymbol{q})$ and $\lambda_\text{im}(\boldsymbol{q})$ are the real and imaginary parts of the eigenvalue respectively, also notice that for each $j$ all partial derivatives are taken for fixed $q_{k \neq j}$. This framework allows us to express the condition of a local maximum with a vanishing value of the real part, $\max_{q_j} \Re(\lambda(\boldsymbol{q})) = 0$, as $\lambda_\text{re}(\boldsymbol{q}) = 0$ and $\frac{\partial \lambda_\text{re}}{\partial q_j}\big|_{\boldsymbol{q}} = 0$ for any parameter $q_j$. Then the equations (S26) can be used to solve for the exact values of $\lambda_\text{im}(\boldsymbol{q})$ and its derivatives $\frac{\partial \lambda_\text{im}}{\partial q_j}\big|_{\boldsymbol{q}}$ at this point, and thus give rise to an implicit differentiation technique for $\lambda_\text{im}(\boldsymbol{q})$. The construction above connects to the more general implicit function theorem [S3].

Below we adapt this approach for finding the threshold value $C_\text{crit}$ and the corresponding $\zeta_{e,\text{crit}}$ by using a system of equations that also solves for $\lambda_\text{im}$ and one of its derivatives. Despite that $\lambda_\text{im}$ is not particularly relevant for us, this approach allows us to get compact analytic expressions without explicitly solving the eigenvalue equation of the $3 \times 3$ matrices $\mathcal{J}^{(2)}$ and $\mathcal{J}_\text{tr}^{(2+3)}$. This allows us to bypass the otherwise required and cumbersome calculus of the Cardano formulas.

### B. Amplification threshold for $\mathcal{J}^{(2)}$

Specializing to the case of $\mathcal{J}^{(2)}$ we identify $q_j = \zeta_e$. The system equations to solve corresponds to the Eq. (S26) corresponding to the real and imaginary parts for $\tilde{\chi}_{\mathcal{J}^{(2)}}(\boldsymbol{q})$ and its $\zeta_e$ derivatives. These equations are solved for the for $\zeta_{e,\text{crit}}$, $C_\text{crit}$, $\lambda_\text{im}(\zeta_{e,\text{crit}}, C_\text{crit}, \bar{\boldsymbol{q}})$ and $\frac{\partial \lambda_\text{im}}{\partial q_j}\big|_{\zeta_{e,\text{crit}}, C_\text{crit}, \bar{\boldsymbol{q}}}$, while $\bar{\boldsymbol{q}} = (\kappa_\mu, \kappa_e, \zeta_\mu)$ are considered fixed. With these

considerations in place, the system of equations reads

$$\Re\left(\chi_{\mathcal{J}^{(2)}}(i\lambda_{\mathrm{im}}(\zeta_{e,\mathrm{crit}}, C_{\mathrm{crit}}, \bar{q}); \zeta_{e,\mathrm{crit}}, C_{\mathrm{crit}}, \bar{q})\right) = 0$$
$$\Im\left(\chi_{\mathcal{J}^{(2)}}(i\lambda_{\mathrm{im}}(\zeta_{e,\mathrm{crit}}, C_{\mathrm{crit}}, \bar{q}); \zeta_{e,\mathrm{crit}}, C_{\mathrm{crit}}, \bar{q})\right) = 0$$
$$\Re\left(\frac{\partial \chi_{\mathcal{J}^{(2)}}}{\partial \lambda}\frac{\partial \lambda}{\partial \zeta_e} + \frac{\partial \chi_{\mathcal{J}^{(2)}}}{\partial \zeta_e}\right)\Bigg|_{\lambda = i\lambda_{\mathrm{im}}(\zeta_{e,\mathrm{crit}}, C_{\mathrm{crit}}, \bar{q}), \frac{\partial \lambda}{\partial q_j}\big|_q = i\frac{\partial \lambda_{\mathrm{im}}}{\partial q_j}\big|_{\zeta_{e,\mathrm{crit}}, C_{\mathrm{crit}}, \bar{q}}, \zeta_{e,\mathrm{crit}}, C_{\mathrm{crit}}, \bar{q}} = 0,\quad\text{(S27)}$$
$$\Im\left(\frac{\partial \chi_{\mathcal{J}^{(2)}}}{\partial \lambda}\frac{\partial \lambda}{\partial \zeta_e} + \frac{\partial \chi_{\mathcal{J}^{(2)}}}{\partial \zeta_e}\right)\Bigg|_{\lambda = i\lambda_{\mathrm{im}}(\zeta_{e,\mathrm{crit}}, C_{\mathrm{crit}}, \bar{q}), \frac{\partial \lambda}{\partial q_j}\big|_q = i\frac{\partial \lambda_{\mathrm{im}}}{\partial q_j}\big|_{\zeta_{e,\mathrm{crit}}, C_{\mathrm{crit}}, \bar{q}}, \zeta_{e,\mathrm{crit}}, C_{\mathrm{crit}}, \bar{q}} = 0,$$

By substituting Eq. (S23a) this gives a non-linear system of four equations for four unknowns

$$\left(\kappa_\mu + \frac{\kappa_e}{2}\right)\lambda_{\mathrm{im}}^2 + \zeta_{e,\mathrm{crit}}\kappa_\mu\lambda_{\mathrm{im}} - \frac{1}{8}\kappa_e\left(\kappa_\mu^2 + 4\zeta_\mu^2\right) = 0, \quad\text{(S28a)}$$
$$\kappa_\mu\left[\lambda_{\mathrm{im}} + (\zeta_{e,\mathrm{crit}} + 2\lambda_{\mathrm{im}})\lambda'_{\mathrm{im}}\right] + \kappa_e\lambda_{\mathrm{im}}\lambda'_{\mathrm{im}} = 0, \quad\text{(S28b)}$$
$$\lambda_{\mathrm{im}}^3 + \zeta_{e,\mathrm{crit}}\lambda_{\mathrm{im}}^2 - \frac{1}{4}\left[\kappa_\mu(\kappa_\mu + 2\kappa_e) + 4\zeta_\mu^2\right]\lambda_{\mathrm{im}} - \frac{1}{4}\zeta_{e,\mathrm{crit}}\left(\kappa_\mu^2 + 4\zeta_\mu^2\right) + \frac{1}{2}\zeta_\mu\kappa_e\kappa_\mu C_{\mathrm{crit}} = 0, \quad\text{(S28c)}$$
$$\lambda_{\mathrm{im}}^2(1 + 3\lambda'_{\mathrm{im}}) + 2\zeta_{e,\mathrm{crit}}\lambda_{\mathrm{im}}\lambda'_{\mathrm{im}} - \frac{1}{4}\kappa_\mu\left[\kappa_\mu + (\kappa_\mu + 2\kappa_e)\lambda'_{\mathrm{im}}\right] - \zeta_\mu^2(1 + \lambda'_{\mathrm{im}}) = 0, \quad\text{(S28d)}$$

where we have introduced the simplified notation $\lambda_{\mathrm{im}} = \lambda_{\mathrm{im}}(\zeta_{e,\mathrm{crit}}, C_{\mathrm{crit}}, \bar{q})$ and $\lambda'_{\mathrm{im}} = \frac{\partial \lambda_{\mathrm{im}}}{\partial \zeta_e}\big|_{\zeta_{e,\mathrm{crit}}, C_{\mathrm{crit}}, \bar{q}}$. The solution of this system follows after some straightforward but cumbersome algebra and by applying the constraint that $\lambda_{\mathrm{im}}$ should be real for self-consistency

$$\lambda_{\mathrm{im}}^{(\pm)} = \pm\frac{1}{2\sqrt{3}}\sqrt{4\zeta_\mu^2 - \kappa_\mu^2 + 2\mathcal{R}}, \quad\text{(S29a)}$$
$$\lambda'_{\mathrm{im}} = -\frac{\kappa_\mu^2 + 4\zeta_\mu^2}{\kappa_\mu(\kappa_\mu + \kappa_e) + 4\zeta_\mu^2 + \frac{\kappa_e}{\kappa_\mu}\mathcal{R}}, \quad\text{(S29b)}$$
$$C_{\mathrm{crit}}^{(\pm)} = \frac{2\lambda_{\mathrm{im}}^{(\pm)}}{3}\frac{\kappa_\mu^2 - 4\zeta_\mu^2 + \mathcal{R}}{\kappa_\mu^2\zeta_\mu}, \quad\text{(S29c)}$$
$$\zeta_{e,\mathrm{crit}}^{(\pm)} = \frac{\lambda_{\mathrm{im}}^{(\pm)}[\kappa_e\mathcal{R} - \kappa_\mu^3 - 4\zeta_\mu^2(\kappa_\mu + \kappa_e)]}{\kappa_\mu(\kappa_\mu^2 + 4\zeta_\mu^2)}, \quad\text{(S29d)}$$

where $\mathcal{R} = \sqrt{\kappa_\mu^4 + 4\zeta_\mu^2\kappa_\mu^2 + 16\zeta_\mu^4}$ and the superscript $\pm$ denotes the solution branch.

These solutions are locally isolated in the $(\lambda_{\mathrm{im}}, \lambda'_{\mathrm{im}}, C_{\mathrm{crit}}, \zeta_{e,\mathrm{crit}})$-space as the Jacobian of the system of Eq. (S23a) is equal to $J^{(\pm)} = \mp\frac{\kappa_e^2\kappa_\mu^2}{2}\lambda_{\mathrm{im}}\mathcal{R}\zeta_\mu$, which is non-zero provided that the detuning $\zeta_\mu$ does not take any of the following values

$$\zeta_\mu \neq 0, \ \zeta_\mu \neq \pm i\frac{\kappa_\mu}{2}, \ \zeta_\mu \neq \pm\frac{1 + i\sqrt{3}}{4}\kappa_\mu, \ \zeta_\mu \neq \pm\frac{1 - i\sqrt{3}}{4}\kappa_\mu. \quad\text{(S30)}$$

This implies that $C_{\mathrm{crit}}$ and $\zeta_{e,\mathrm{crit}}$ can be thought as single-valued functions of $\zeta_\mu$ for fixed $\kappa_e, \kappa_\mu$ as long as $\zeta_\mu$ does not cross zero (the other solutions are complex and thus irrelevant for $\zeta_\mu \in \mathbb{R}$). This behavior is physical as the $(+)$ branch should be chosen for $\zeta_\mu > 0$ and the $(-)$ one should be chosen for $\zeta_\mu < 0$, in order to ensure that $C_{\mathrm{crit}} > 0$ required by Eq. (S25). The above enable us to minimize $C_{\mathrm{crit}}$ in terms of $\zeta_\mu$ leading to the well-known result $\min_{\zeta_\mu} C_{\mathrm{crit}} = 1$ for $\zeta_\mu \to \pm\infty$. This corresponds to the solutions for $\mathcal{J}^{(2)}$ depicted in Fig. 3(a) and 3(b) of the main text, the absence of a superscript is due to the branch fixing mentioned above.

Notice also that all the above results for $C_{\mathrm{crit}}$ are consistent with Ref. [S2]. This groundwork allows us to generalize to the $g_3 \neq 0$ case without any substantial further calculation.

### C. Amplification threshold for $\mathcal{J}_{\mathrm{tr}}^{(2+3)}$

To obtain the corresponding set of solutions for the $\mathcal{J}_{\mathrm{tr}}^{(2+3)}$ dynamical matrix we perform the transformation of cooperativity and detuning noted in Eq. (S23b). This directly leads to the modification of the solution of Eq. (S29c)

and (S29d) to

$$C_{\text{crit}}^{(\pm)} = \frac{2\lambda_{\text{im}}^{(\pm)}}{3} \frac{\kappa_\mu^2 - 4\Delta^2 + \mathcal{R}}{\kappa_\mu^2 \Delta_+}, \tag{S31a}$$

$$\zeta_{e,\text{crit}}^{(\pm)} = \frac{\lambda_{\text{im}}^{(\pm)}[\kappa_e \mathcal{R} - \kappa_\mu^3 - 4\Delta^2(\kappa_\mu + \kappa_e)]}{\kappa_\mu(\kappa_\mu^2 + 4\Delta^2)}, \tag{S31b}$$

where $\lambda_{\text{im}}^{(\pm)}$ and $\lambda'_{\text{im}}$ are equivalent to Eq. (S29a) and (S29b) respectively after the substitution $\zeta_\mu \to \Delta$.

Similarly as before these solutions are locally isolated in the $(\lambda_{\text{im}}, \lambda'_{\text{im}}, C_{\text{crit}}, \zeta_{e,\text{crit}})$-space as the Jacobian of the system of Eq. (S23b) is equal to $J^{(\pm)} = \mp \frac{\kappa_e^2 \kappa_\mu^2}{2} \lambda_{\text{im}} \mathcal{R} \Delta_+$, which is non-zero provided that the detuning $\zeta_\mu$ does not take any of the following values

$$\zeta_\mu \neq g_3|a_0|^2, \qquad \zeta_\mu \neq 2g_3|a_0|^2 \pm \sqrt{(g_3|a_0|^2)^2 - \frac{\kappa_\mu^2}{4}},$$

$$\zeta_\mu \neq 2g_3|a_0|^2 \pm \sqrt{(g_3|a_0|^2)^2 - \left(\frac{1+i\sqrt{3}}{2}\right)\frac{\kappa_\mu^2}{4}}, \quad \zeta_\mu \neq 2g_3|a_0|^2 \pm \sqrt{(g_3|a_0|^2)^2 - \left(\frac{1-i\sqrt{3}}{2}\right)\frac{\kappa_\mu^2}{4}}, \tag{S32}$$

This shows that branch change in the solution can exist at the point where $C_{\text{crit}}$ diverges, i.e. at $\Delta_+ = 0$, and at the points where Kerr amplification occurs, see also Eq. (S20). The other two points are complex and thus irrelevant since $\zeta_\mu \in \mathbb{R}$. In practice since we restrict ourselves to the $g_3|a_0|^2 \leq \frac{\kappa_\mu}{4}$ regime where Kerr-amplification is impossible the branch fixing gives that the $(+)$ branch is selected for $\Delta_+ > 0$ and the $(-)$ branch is selected for $\Delta_+ < 0$. With the branch fixing in place these results are also depicted in the Fig. 3(a) and 3(b) of the main text.

Finally, let us note that in the main text we also show in Fig. 3(c)–3(d) the minimum values of $C_{\text{crit}}$, $\zeta_\mu$ and $\zeta_{e,\text{crit}}$ as a function of $g_3|a_0|^2/\kappa_\mu$. These stem from a simple minimization of Eq. (S31a), yielding the equations Eq. (S6), (S7) and (S8). Since this process is straightforward here we refrain from elaborating further.

## IV. TIME-DOMAIN VERIFICATION VIA SECOND-MOMENT PROPAGATION

To verify the onset criterion derived from the eigenvalues of the linearized drift matrix, we simulate the *linearized* quantum Langevin dynamics in the time domain at the level of *second moments*. This avoids stochastic trajectory averaging and directly captures exponential growth when the linear system is unstable.

### A. Linearized operator vector and drift matrix

We work with the six-component operator vector

$$u \equiv \left(a_+, a_+^\dagger, a_-, a_-^\dagger, b, b^\dagger\right)^\mathsf{T}, \tag{S33}$$

where $a_\pm \equiv a_{\pm 1}$ are the optical sidebands and $b$ is the microwave mode. Linearizing around the pumped steady amplitude $a_0$ yields

$$\dot{u}(t) = M\, u(t) + K\, \xi(t), \tag{S34}$$

with a constant $6 \times 6$ drift matrix $M$ and a coupling matrix $K$ that injects vacuum/thermal noise from internal and external ports.

In the ordering (S33), the drift matrix used in the simulations is

$$M = \begin{pmatrix} -\frac{\kappa_\mu}{2} - i\zeta_\mu + 2ig_3|a_0|^2 & 0 & 0 & ig_3 a_0^2 & -ig_2 a_0 & 0 \\ 0 & -\frac{\kappa_\mu}{2} + i\zeta_\mu - 2ig_3|a_0|^2 & -ig_3 a_0^{*2} & 0 & 0 & ig_2 a_0^* \\ 0 & ig_3 a_0^2 & -\frac{\kappa_\mu}{2} - i\zeta_\mu + 2ig_3|a_0|^2 & 0 & 0 & -ig_2 a_0 \\ -ig_3 a_0^{*2} & 0 & 0 & -\frac{\kappa_\mu}{2} + i\zeta_\mu - 2ig_3|a_0|^2 & ig_2 a_0^* & 0 \\ -ig_2 a_0^* & 0 & 0 & -ig_2 a_0 & -\frac{\kappa_e}{2} - i\zeta_e & 0 \\ 0 & ig_2 a_0 & ig_2 a_0^* & 0 & 0 & -\frac{\kappa_e}{2} + i\zeta_e \end{pmatrix}. \tag{S35}$$

This is the time-domain drift corresponding to the same linearization used for the eigenvalue threshold: if $\max \Re \lambda(M) > 0$, the linearized dynamics is unstable and second moments grow exponentially.

## B. Noise ports and diffusion matrix

We include independent internal and external noise inputs for each optical sideband and for the microwave mode. We collect them in the 12-component vector

$$\xi \equiv \left(A_{+,\text{in}},\, A^\dagger_{+,\text{in}},\, A_{+,\text{ex}},\, A^\dagger_{+,\text{ex}},\, A_{-,\text{in}},\, A^\dagger_{-,\text{in}},\, A_{-,\text{ex}},\, A^\dagger_{-,\text{ex}},\, B_{\text{in}},\, B^\dagger_{\text{in}},\, B_{\text{ex}},\, B^\dagger_{\text{ex}}\right)^\mathsf{T}. \tag{S36}$$

The Markov correlations are encoded as

$$\langle \xi(t)\, \xi^\mathsf{T}(t')\rangle = N\, \delta(t-t'), \tag{S37}$$

where $N$ is a constant $12 \times 12$ matrix whose nonzero $2 \times 2$ blocks implement, for each bath with occupancy $\bar{n}$,

$$\langle C(t)\, C^\dagger(t')\rangle = (\bar{n}+1)\delta(t-t'), \qquad \langle C^\dagger(t)\, C(t')\rangle = \bar{n}\,\delta(t-t'). \tag{S38}$$

In our parameter choices the optical baths are taken as vacuum ($\bar{n} \simeq 0$) while the microwave internal bath may be thermal ($\bar{n}_{e,\text{in}} \neq 0$).

The coupling matrix $K$ injects these noises into the six intracavity operators. Writing $\kappa_{o,\text{ex}} = \eta_o \kappa_o$, $\kappa_{o,\text{in}} = (1-\eta_o)\kappa_o$ and similarly for $\kappa_e$, we use

$$D \equiv KNK^\mathsf{T}, \tag{S39}$$

as the diffusion matrix appearing in the second-moment equation below. (This transpose convention matches the correlator definition $\langle \xi \xi^\mathsf{T}\rangle$ in (S37).)

## C. Second-moment (Lyapunov) equation and observables

Define the $6 \times 6$ second-moment matrix

$$R(t) \equiv \langle u(t)\, u^\mathsf{T}(t)\rangle. \tag{S40}$$

Using (S34) and the Markov property (S37), $R(t)$ obeys the closed linear matrix ODE

$$\dot R = MR + RM^\mathsf{T} + D, \tag{S41}$$

which we integrate numerically with $R(0) = 0$ (vacuum/thermal initial condition for the intracavity fluctuations).

With the ordering (S33), the mode populations are read off directly:

$$n_+(t) = \langle a^\dagger_+ a_+ \rangle = \Re R_{2,1}(t), \qquad n_-(t) = \langle a^\dagger_- a_- \rangle = \Re R_{4,3}(t), \qquad n_b(t) = \langle b^\dagger b\rangle = \Re R_{6,5}(t). \tag{S42}$$

## D. Simulation scenarios

Figure 4 compares three operating points that isolate the roles of the $\chi^{(2)}$ and Kerr ($\chi^{(3)}$) nonlinearities. Throughout we use the parameters quoted in the main text: $\kappa_\mu/2\pi = 11$ MHz, $\kappa_e/2\pi = 1.1$ MHz, $\eta_\mu = \eta_e = 0.2$, $\zeta_\mu/2\pi = 11$ MHz, $\zeta_e/2\pi = 0$. We then consider: (i) $\chi^{(2)}$ only: $g_3 = 0$, $C = 0.1$, (ii) Kerr only: $C = 0$ with $g_3|a_0|^2/\kappa_\mu = 0.49$, (iii) combined $\chi^{(2)} + \chi^{(3)}$: the same subthreshold $C = 0.1$ together with a Kerr strength $g_3|a_0|^2/\kappa_\mu = 0.49$. In this case $\max \Re \lambda(M) > 0$ and a masing instability is observed.

As explicitly shown and discussed in the main text in cases (i) and (ii) Eq. (S41) relaxes to a bounded steady state. In contrast, in case (iii) $\max \Re \lambda(M) > 0$ and the second moments $n_+(t)$ and $n_b(t)$ grow exponentially. This confirms dynamically that the combined nonlinearities can cross threshold even when each nonlinearity alone is subthreshold.

Finally, we note that the employed parameters are realistic. For a pump power $P_{\text{in}} = 20$ mW at $\omega_0/2\pi = 193$ THz resulting to a population $|a_0|^2 \approx \frac{4\eta_\mu P_{\text{in}}}{\kappa_\mu \hbar \omega_0}$, $C = 0.1$ corresponds to $g_2/2\pi \approx 12.93$ Hz and $g_3|a_0|^2/\kappa_\mu = 0.49$ to $g_3/2\pi \approx 2.98 \times 10^{-3}$ Hz, close to experimentally achieved values [S4].



\end{document}